# A Mathematical Reformulation of the Reference Price

by

Kevin D. Dayaratna

P. K. Kannan[*]

May 22, 2012

[*] Kevin D. Dayaratna is a doctoral student in the Department of Mathematics, University of Maryland, and P. K. Kannan is Ralph J. Tyser Professor of Marketing Science, Smith School of Business, University of Maryland. All correspondence should be addressed to Kevin D. Dayaratna at kevind@math.umd.edu. We would like to thank Hovannes Abramyan, Erik Johannassen, Steven J. Miller, Praveen Kopalle, Brian Ratchford, Mike Taylor, and Chi Kin (Bennett) Yim for helpful comments on an earlier draft of this manuscript.

# A Mathematical Reformulation of the Reference Price

**Abstract**

Reference prices have long been studied in applied economics and business research. One of the classic formulations of the reference price is in terms of an iterative function of past prices. There are a number of limitations of such a formulation, however. Such limitations include burdensome computational time to estimate parameters, an inability to truly account for customer heterogeneity, and an estimation procedure that implies a misspecified model. Managerial recommendations based on inferences from such a model can be quite misleading. We mathematically reformulate the reference price by developing a closed-form expansion that addresses the aforementioned issues, enabling one to elicit truly meaningful managerial advice from the model. We estimate our model on a real world data set to illustrate the efficacy of our approach. Our work is not only useful from a modeling perspective, but also has important behavioral and managerial implications, which modelers and non-modelers alike would find useful.

# I. Introduction

For decades, it has been well understood that people use benchmarks in making purchasing decisions. For example, if a customer at a grocery store sees several brands of a particular product, her choice about which brand to purchase will depend on a variety of factors. Such factors may include how much she has paid for these brands in the past as well as her overall experiences having tried those brands before (Hardie, Johnson, and Fader 1993).

Marketing researchers have used the concept of the "reference price" to quantify how customers incorporate price stimuli experienced during purchase occasions. Some research has modeled reference prices, referring to them as *stimulus-based reference prices,* based on information available at the time of the purchase (Hardie, Johnson, and Fader 1993, Rajendran and Tellis 1994, Mazumdar and Paptala 1995). Other work has modeled the reference price as a weighted average of past prices encountered with varying carry-over weights, appropriately called *memory-based reference prices* (Lattin and Bucklin 1989, Kalyanaraman and Little 1994; Mazumdar and Papatla 1995). There has also been some work incorporating explanatory variables in addition to past prices, such as price trends and market share (Winer 1986; Kalwani et al 1990, Kalwani and Yim 1992, Kopalle et al 1996). Such alternative formulations of reference prices are usually the first step in using them as exogenous variables in choice models, where choice of product alternatives is explained as a function of their prices and the deviation of those prices from the reference price. If the price of a brand during a purchase occasion is less than the customer's reference price, the deviation is termed as a "gain" and is otherwise termed as a "loss." Customer choice is then modeled as a function of brand prices, gains, and losses. Briesch et al. (1997) provides a review of such reference price formulations and their use in choice models based on extant studies.

Even in recent years, there has been a plethora of work examining reference pricing. Mazumdar, Raj, and Sinha (2005), Liu and Soman (2008), and Heyman and Mellers (2008) review some of this research. For example, Heath et al (1995) examine reference dependence from the perspective of mental accounting. Janiszewski and Lichtenstein (1999) apply Range Theory to the perception of reference prices and determined that the variance in the width of the price range affects consumer attitudes pertaining to the attractiveness of the price. Kamins et al (2004) looks at reference pricing in terms of internet auctions. Cunha and Shulman (2011) studies how information processing goals can affect price judgments with respect to other products. More recently, Kopalle et al (2012) show how heterogeneity in household reference prices can be an important input to optimal pricing policies.

In this research, we focus on the formulation of memory-based reference prices. Existing models that define the reference price as a weighted average of past prices with varying carry-over weights have a few notable limitations. First, since the reference prices are used exogenously in choice models, they are calibrated first (outside the choice model) on a subset of the data by using a grid search on carryover weights corresponding to how customers weigh past prices. This brute-force method does not use all information available in the sample, as only a subset of the data is used for estimation. Secondly, this exogenous estimation of weights results in coefficient estimates of a misspecified model. Thirdly, these models assume that all customers weight past and present prices in the same manner, ignoring customer heterogeneity. Our research presents a reformulation that overcomes these problems and consequently has important managerial and behavioral implications.

In the next section, we discuss the problems with existing memory-based reference price models. We then present a new model that addresses these limitations and estimate our model on a real world data set. Lastly, we discuss our results and conclude with potential avenues of future research.

**II. Current Formulation of the Reference Price and Associated Problems**

A commonly-used operationalization of the reference price, called the memory-based reference price, defines the reference price of brand $j$ at time $t$ in terms of a brand's price and its reference price at the preceding time $t$-1 (Kalwani and Yim 1992; Lattin and Bucklin 1989; Kalyanaraman and Little 1994; Mazumdar and Papatla 1995). Specifically, if $r_{j,t-1}$ is the reference price for brand $j$ at time $t$-1 and $p_{j,t-1}$ is actual price at time $t$-1, the reference price at time $t$, $r_{j,t}$, can be defined as follows:

$$r_{j,t} = \pi \cdot r_{j,t-1} + (1-\pi) \cdot p_{j,t-1}.$$

The coefficient $\pi$ corresponds to the carry-over weights attached to past reference prices and actual prices.

Reference price defined in such a way has been included exogenously as an explanatory variable in choice models (typically as "gain" and "loss" variables) along with regular price and other explanatory variables. Since these reference prices have to be determined before including them as explanatory variables and are defined iteratively, researchers would typically estimate the $\pi$ coefficients exogenously via a grid search and use those estimates to compute the reference price over the time series. Typically, a portion of the total sample is set aside (say the time-series data covering Time Period *1* through *T*) as an "initialization sample" which is used for estimating the carry-over weight. The reference prices are then determined based on these estimated weights and used in the remaining portion of the data (covering Time Period *T* onwards) to estimate the model.

There are several problems with this approach. First, an accurate grid search is burdensome computationally as it can take a significant amount of time to estimate the model. Additionally, dividing the total observations as an initialization sample for the grid search and a calibration sample for the choice model does not allow the use of all observations for estimating the carry-over weights and the parameters of the choice model.

Thirdly, the approach assumes the same values of $\pi$ for *all* customers, ignoring customer heterogeneity. Assuming that all customers weigh past prices in the same manner is not realistic because more frequent purchases may lead to prices being more accurately remembered (see Briesh et al. 1997). Thus, different customers will almost surely vary in their use of reference prices.

Lastly, and most importantly, such an approach is statistically unsound. By estimating the values of $\pi$ separately via a grid search and then estimating the choice model, the resulting coefficient estimates are of a misspecified model. We can understand why via the following theoretical argument. Without loss of generality, suppose that we have two models $P_\Theta$ and $P_{\Theta_0}$, with densities $f(x;\vartheta)$ and $f(x;\vartheta_0)$ with respect to Lebesgue measure. Let $P_\Theta$ be our true model and $P_{\Theta_0}$ be the model that we specify. The Kullback-Leibler divergence, $K(\vartheta,\vartheta_0)$ is defined as:

$$K(\vartheta_0, \vartheta) = -E_{\vartheta_0} \log\left(\frac{f(x;\vartheta)}{f(x;\vartheta_0)}\right) = -\int_{-\infty}^{\infty} \log\left(\frac{f(x;\vartheta)}{f(x;\vartheta_0)}\right) f(x;\vartheta_0) dx.$$

As the maximum likelihood estimator minimizes the Kullback-Leibler divergence (Van der Vaart 2007), the previously discussed two-step estimation procedure results in a misspecified model. This misspecification is due to the researcher is estimating a conditional choice model *given* $\pi$ rather than actually incorporating $\pi$ into the model. Consequently, the coefficient estimates do not truly reflect reality.

We address these problems in this research. Specifically, we reformulate the memory-based reference price in terms of a closed-form expansion so that it can be incorporated directly into the choice model and carry-over weight estimated simultaneously with the other parameters of the choice model to resolve the previously discussed issues of model misspecification. This reformulation also allows for the incorporation of heterogeneity in carry-over weights in the choice model and enables estimation of potentially different weights for different groups of customers using finite mixture estimation methods (Kamakura and Russell, 1989).

### III. Reformulating the Reference Price

Recall the iterative definition of the reference price $r_{j,t}$ for brand *j* and time *t* :

$$r_{j,t} = \pi \cdot r_{j,t-1} + (1-\pi) \cdot p_{j,t-1} \quad (*).$$

where $p_{j,t-1}$ is the brand's price in the previous time period. We claim, and prove by mathematical induction, that an equivalent closed-form, non-iterative representation of the reference price can be defined as follows:

$$r_{j,t} = \pi^{t-1} \cdot p_{j,1} + (1-\pi)\sum_{i=1}^{t-1} \cdot \pi^{i-1} p_{j,t-i} \quad (**).$$

Proof (by induction):

For *t=1:*

$r_{j,1} = p_{j,1}$, which coincides with the current formulation.

We assume our inductive hypothesis by assuming our expression holds at time *t* and show that our expression holds for time *t+1:*

$$r_{j,t} = \pi^{t-1} \cdot p_{j,1} + (1-\pi)\sum_{i=1}^{t-1} \cdot \pi^{i-1} p_{j,t-i}.$$

Substituting the above into (*) gives:

$$r_{j,t+1} = \pi \cdot r_{j,t} + (1-\pi) \cdot p_{j,t}$$

$$= \pi \cdot \left( \pi^{t-1} \cdot p_{j,1} + (1-\pi)\sum_{i=1}^{t-1} \cdot \pi^{i-1} p_{j,t-i} \right) + (1-\pi) \cdot p_{j,t}$$

$$= \pi^{t} \cdot p_{j,1} + (1-\pi)\sum_{i=1}^{t-1} \cdot \pi^{i} p_{j,t-i} + (1-\pi) \cdot p_{j,t}$$

$$= \pi^{t} \cdot p_{j,1} + (1-\pi)\sum_{i=0}^{t-1} \cdot \pi^{i} p_{j,t-i}$$

$$r_{j,t+1} = \pi^{t} \cdot p_{j,1} + (1-\pi)\sum_{i=1}^{t} \cdot \pi^{i-1} p_{j,(t+1)-i}.$$

Q.E.D.

This is our closed-form representation of the reference price. In the next section we incorporate this definition into an integrated purchase choice/incidence model to analyze customer behavior.

## IV. Choice Model Development and Estimation

We now discuss our choice/incidence model incorporating reference price as an explanatory variable. Behavioral theory suggests that people view negative changes more seriously than positive changes (Kahenman and Tversky 1979). Quantitatively, one can incorporate this concept into a utility function for a customer examining brand $j$ at time $t$ (Kalwani et al 1990; Kopalle et al 1996):

$$u_{j,t} = \beta_j + \beta_p p_{j,t} + \begin{cases} (r_{j,t} - p_{j,t})\beta_g & if\ r_{j,t} > p_{j,t} \\ (r_{j,t} - p_{j,t})\beta_l & if\ r_{j,t} \leq p_{j,t} \end{cases}$$

where $\beta_j$ is an intercept parameter for each brand, $\beta_p$ is a coefficient corresponding to price, $\beta_g$ is a coefficient (that we refer to as a gain parameter) corresponding to how the consumer views a reference price that is greater than the actual price, and $\beta_l$ is a coefficient (that we refer to as a loss parameter) corresponding to how the consumer views a reference price that is less than the actual price.

Armed with our new closed-form expansion, we can rewrite our utility function as follows:

$$u_{j,t} = \beta_j + \beta_p p_{j,t} + \begin{cases} \left(\pi^{t-1} \cdot p_{j,1} + (1-\pi)\sum_{i=1}^{t-1} \cdot \pi^{i-1} p_{j,t-i} - p_{j,t}\right)\beta_g & \text{if } r_{j,t} > p_{j,t} \\ \left(\pi^{t-1} \cdot p_{j,1} + (1-\pi)\sum_{i=1}^{t-1} \cdot \pi^{i-1} p_{j,t-i} - p_{j,t}\right)\beta_l & \text{if } r_{j,t} \leq p_{j,t} \end{cases}.$$

We can go can go a step further and use the Heaviside step function to avoid having to impose the above inequality restrictions (see Appendix for further detail).

Now that we have a closed-form utility function incorporating the reference price, we can use our function to estimate a choice/incidence model using a nested logit formulation (Kannan and Wright 1991; Bucklin and Gupta 1992). The probability an individual chooses brand $j$ in category $B$, $\Pr_t(j \cap B)$, is:

$$\Pr_t(j \cap B) = \Pr_t(j|B)\Pr_t(B)$$
$$= \frac{\exp(u_{j,t})}{\sum_{k=1}^{K}\exp(u_{k,t})} \frac{\exp(\alpha_0 + \alpha_1 CV_t)}{1 + \exp(\alpha_0 + \alpha_1 CV_t)}$$

$$\Pr_t(j \cap B) = \frac{\exp(u_{j,t})}{\sum_{k=1}^{K}\exp(u_{k,t})}\left(\frac{\exp\left(\alpha_0 + \alpha_1 \log\left(\sum_{k=1}^{K}\exp(u_{k,t})\right)\right)}{1 + \exp\left(\alpha_0 + \alpha_1 \log\left(\sum_{k=1}^{K}\exp(u_{k,t})\right)\right)}\right)$$

where $\alpha_0$ is an intercept parameter and $\alpha_1$ is a coefficient corresponding to category value. Assuming latent segments where $\psi_s$ is the fraction of customers that belong to a particular segment $s$, we have:

$$\Pr{}_t(j \cap B) = \sum_{s=1}^{S} \psi_s \frac{\exp(u_{j,t,s})}{\sum_{k=1}^{K} \exp(u_{k,t,s})} \left( \frac{\exp\left(\alpha_{0,s} + \alpha_{1,s} \log\left(\sum_{k=1}^{K} \exp(u_{k,t,s})\right)\right)}{1 + \exp\left(\alpha_{0,s} + \alpha_{1,s} \log\left(\sum_{k=1}^{K} \exp(u_{k,t,s})\right)\right)} \right).$$

In the next section, we present an empirical illustration of our methodology in estimating model coefficients for a finite mixture three segment model.

## V. Empirical Illustration

We estimated our model using an IRI scanner-panel cola data set on household purchases from a large U.S. city (Chib et al, 2004). Our data set was collected between 1991 and 1993, containing information about 350 customers and 104 observations. There were four brands in the cola category in these data, and we set the alternative specific constant for one brand to zero as a benchmark to ensure statistical identifiability of our model.

We estimated our coefficients via maximum likelihood estimation for a three segment latent class model. Results are below. (*) indicates that the coefficient estimate is statistically significant below a critical threshold of 0.01.

Three segment model: Reformulation

| Coefficient | Parameter | Segment 1 | Segment 2 | Segment 3 |
|---|---|---|---|---|
| Reference Price Weight | $\pi_s$ | 0.0770 (*) | 0.2894 (*) | 0.6628 (*) |
| Incidence | $\alpha_{0,s}$ | 5.7656 (*) | 0.5000 | 1.0497 (*) |

| Coefficient | Parameter | Segment 1 | Segment 2 | Segment 3 |
|---|---|---|---|---|
| Intercept | | | | |
| Category Value | $\alpha_{1,s}$ | 3.9185 (*) | 0.7858 (*) | 7.0767 (*) |
| Brand 1 Intercept | $\beta_{1,s}$ | 3.2919 (*) | 1.0472 (*) | 0.7575 (*) |
| Brand 2 Intercept | $\beta_{2,s}$ | 6.5073 (*) | 0.6386 (*) | 1.8804 (*) |
| Brand 3 Intercept | $\beta_{3,s}$ | 6.2598 (*) | 0.6493 (*) | 0.7543 (*) |
| Gain | $\beta_{g,s}$ | 0.8500 (*) | 0.6966 (*) | 0.7511 (*) |
| Loss | $\beta_{l,s}$ | 1.1160 (*) | 7.4963 (*) | 6.9011 (*) |
| Price | $\beta_{p,s}$ | -4.8143 (*) | -10.5571 (*) | -9.8313 (*) |
| Segment Size | $\psi_s$ | 0.0905 (*) | 0.4444 (*) | 0.4651 (*) |

Log-likelihood: -56104.72

We also estimated our model using the existing two step method:

Three Segment Model: Existing Two Step Estimation Procedure

| Coefficient | Parameter | Segment 1 | Segment 2 | Segment 3 |
|---|---|---|---|---|
| Reference Price Weight | $\pi$ | | 0.8315 (*) | |
| Incidence | $\alpha_{0,s}$ | 0.6036 (*) | 6.6575 (*) | 1.0151 |

| Intercept | | | | |
|---|---|---|---|---|
| Category Value | $\alpha_{1,s}$ | 7.1112 | 2.5017 | 1.2982 |
| Brand 1 Intercept | $\beta_{1,s}$ | 0.6315 (*) | 4.3144 (*) | 0.7030 (*) |
| Brand 2 Intercept | $\beta_{2,s}$ | 2.2023 (*) | 7.5313 (*) | 1.2950 (*) |
| Brand 3 Intercept | $\beta_{3,s}$ | 1.2775 (*) | 6.9941 (*) | 1.4746 (*) |
| Gain | $\beta_{g,s}$ | 1.6437 (*) | 1.3524 (*) | 7.3583 (*) |
| Loss | $\beta_{l,s}$ | 11.1287 (*) | 11.6966 (*) | 11.0801 |
| Price | $\beta_{p,s}$ | -9.6259 (*) | -4.4994 (*) | -11.3123 (*) |
| Segment Size | $\psi_s$ | 0.2430 (*) | 0.1053 (*) | 0.6517 (*) |

Log-likelihood: -32656.34

In both models, virtually all of our coefficient estimates are highly significant, including our loss parameters which are larger than our gain parameters (Kahneman and Tversky 1979). The two step procedure, however, only provides us with one value of $\pi$ for all segments (0.8315), whereas we had separate values for our new procedure. In fact, our statistically significant estimates of $\pi_s$ in our three segment reformulated model implies that customers are quite heterogeneous in making purchasing decisions. Results from our procedure, which takes into account information from all data, suggest that the overall carryover is significantly inflated in the two-step procedure. Specifically, the estimate of $\pi$ from the two-step procedure implies a much longer-term memory of prices than our estimates across the segments indicate and could lead to incorrect managerial decisions with regard to pricing and promotions (cf., Kopalle et al 2012).

From a computational perspective, our model is also much easier to implement than the existing procedure. Both model estimations were performed using MATLAB on a 2.2 GHZ AMD processor with 4 GB of RAM, a quite modest machine by today's standards. Our estimation took nearly twenty four times longer for the existing two step estimation procedure compared to our method (12 hours versus 35 minutes) and would take even longer when choosing finer grids from which to estimate $\pi$. These computational gains are extremely important for large data sets.

Additionally, our reformulation used the entire data set for the estimation, unlike the existing two-step estimation procedure which uses an initialization sample to estimate $\pi$ and uses the remaining fraction of the data to estimate the remaining coefficients. Any frequentist statistical estimation conducted in such a manner becomes particularly problematic with small/medium sized data sets where the asymptotic theory that constitutes the basis for typical maximum likelihood estimation becomes questionable. Resulting maximum likelihood estimates are therefore not useful in providing meaningful managerial advice.

Altogether, our results based on our reformulation provide correct estimates from which managers can make inferences. Additionally, our model enables researchers to properly quantify the long-standing behavioral concept of prospect theory (Kahneman and Tversky 1979). We discuss these ideas in greater detail in the following section.

## VI. Conclusions

We have presented a closed-form reformulation of the memory-based reference price that incorporates customer heterogeneity. Our coefficient estimates are highly significant, signifying that the explanatory variables included in our choice/incidence model explain our situation quite well. Furthermore, our results verify long-standing behavioral theory that losses loom larger than gains.

Although heavily quantitative, we believe that our work proposes substantive implications for non-modelers:

- Statistically speaking, coefficient estimates from our reformulation are much more meaningful now that our model is properly specified. Therefore, managerial recommendations based on our reformulation are not only correct but also much more insightful.
- Reference prices in choice models often attempt to quantify prospect theory (Kahneman and Tversky, 1979) by modeling how consumers react to differences between actual prices and reference prices via gain and loss parameters. Again, as the previously, two step estimation procedure implies a misspecified model, resulting gain and loss parameter estimates do not properly quantify these reactions. Our reformulation assuages this issue.
- There is no a priori reason that our model should only apply to frequently purchased items such as Cola. Other product categories where customers remember recent purchases including durable goods as books (see, Jank and Kannan 2005), or cellular phones (where prices may be are advertised heavily) and services such as cellular phone, internet, and cable television plans (Grewal et al 2011) are just as applicable.

We hope our reformulation encourages future research in statistically modeling reference prices. One potentially fruitful avenue of future research could be to use Bayesian statistics to model individual level customer heterogeneity and discuss the resulting managerial implications. Secondly, it could be interesting to combine a Bayesian approach with the finite mixture approach presented here and draw customer-level inferences within each latent segment.

# VIII. Technical Appendix with Mathematical Details

With our utility function for brand j at time t:

$$u_{j,t} = \beta_j + \beta_p p_{j,t} + \begin{cases} \left(\pi^{t-1} \cdot p_{j,1} + (1-\pi)\sum_{i=1}^{t-1} \pi^{i-1} p_{j,t-i} - p_{j,t}\right)\beta_g & \text{if } r_{j,t} > p_{j,t} \\ \left(\pi^{t-1} \cdot p_{j,1} + (1-\pi)\sum_{i=1}^{t-1} \pi^{i-1} p_{j,t-i} - p_{j,t}\right)\beta_l & \text{if } r_{j,t} \leq p_{j,t} \end{cases}$$

we can go can go a step further and use the Heaviside step function $H$ to avoid having to impose the above inequality restrictions:

$$u_{j,t} = \beta_j + \beta_p p_{j,t} + \left(\pi^{t-1} \cdot p_{j,1} + (1-\pi)\sum_{i=1}^{t-1} \pi^{i-1} p_{j,t-i} - p_{jt}\right)\beta_g^{H\left(\pi^{t-1} \cdot p_{j,1} + (1-\pi)\sum_{i=1}^{t-1} \pi^{i-1} p_{j,t-i} - p_{j,t} - \varepsilon\right)} \cdot$$

$$\beta_l^{H\left(-\left(\pi^{t-1} \cdot p_{j,1} + (1-\pi)\sum_{i=1}^{t-1} \pi^{i-1} p_{j,t-i} - p_{j,t}\right)\right)}.$$

with an arbitrarily small $\varepsilon \sim 0.000001$ to accommodate for the inequality constraints. The choice of $\varepsilon$ does not matter as long as it is non-zero within machine precision.

Now that we have a closed-form utility function incorporating the reference price, we can use our function to estimate a model of purchase choice and incidence using a nested logit formulation (Kannan and Wright 1991; Bucklin and Gupta 1992). The probability that an individual chooses brand $j$ in category $B$, $\Pr_t(j \cap B)$, is:

$$\Pr_t(j \cap B) = \Pr_t(j | B) \Pr_t(B)$$
$$= \frac{\exp(u_{j,t})}{\sum_{k=1}^{K} \exp(u_{k,t})} \frac{\exp(\alpha_0 + \alpha_1 CV_t)}{1 + \exp(\alpha_0 + \alpha_1 CV_t)}$$

$$\Pr_t(j \cap B) = \frac{\exp(u_{j,t})}{\sum_{k=1}^{K} \exp(u_{k,t})} \left( \frac{\exp\left(\alpha_0 + \alpha_1 \log\left(\sum_{k=1}^{K} \exp(u_{k,t})\right)\right)}{1 + \exp\left(\alpha_0 + \alpha_1 \log\left(\sum_{k=1}^{K} \exp(u_{k,t})\right)\right)} \right)$$

where $\alpha_0$ is an intercept parameter and $\alpha_1$ is a coefficient corresponding to category value. Assuming latent segments where $\psi_s$ is the fraction of customers that belong to a particular segment *s*, in our finite mixture model formulation we have:

$$\Pr_t(j \cap B) = \sum_{s=1}^{S} \psi_s \frac{\exp(u_{j,t,s})}{\sum_{k=1}^{K} \exp(u_{k,t,s})} \left( \frac{\exp\left(\alpha_{0,s} + \alpha_{1,s} \log\left(\sum_{k=1}^{K} \exp(u_{k,t,s})\right)\right)}{1 + \exp\left(\alpha_{0,s} + \alpha_{1,s} \log\left(\sum_{k=1}^{K} \exp(u_{k,t,s})\right)\right)} \right).$$

Therefore, our likelihood function, which we conduct maximum likelihood estimation on, is:

$$L(j) = \prod_{i=1}^{I} \prod_{j=1}^{J} \prod_{t=1}^{T} \sum_{s=1}^{S} \left( \psi_s \left( \frac{\exp(u_{j,t,s})}{1 + \sum_{k=1}^{K-1} \exp(u_{k,t,s})} \right)^{y_{i,j,t}} \left( \frac{\exp\left(\alpha_{0,s} + \alpha_{1,s} \log\left(\sum_{k=1}^{K-1} \exp(u_{k,t,s})\right)\right)}{1 + \exp\left(\alpha_{0,s} + \alpha_{1,s} \log\left(\sum_{k=1}^{K-1} \exp(u_{k,t,s})\right)\right)} \right)^{x_{i,t}} \left( \frac{1}{1 + \exp\left(\alpha_{0,s} + \alpha_{1,s} \log\left(\sum_{k=1}^{K-1} \exp(u_{k,t,s})\right)\right)} \right) \right).$$

where $y_{ijt} = 1$ if household *i* brand *j* is bought at time *t* and is 0 otherwise and $x_{i,t} = \sum_{j=1}^{J} y_{i,j,t}$.